\documentclass[aps,prd,superscriptaddress,preprint,tightenlines,nofootinbib,floatfix]{revtex4}

\usepackage{graphicx} % Include figure files
\usepackage{dcolumn}  % Align table columns on decimal point
\usepackage{bm}       % bold math

\usepackage{latexsym}
\usepackage{amsmath}
\usepackage{xspace}

\begin{document}

%\preprint line(s) will be ignored for PRL/PRD
%\preprint{CLEO Draft YY-NNA} % For paper draft CBX YY-NN -> Draft YY-NNA
%\preprint{CLEO CONF YY-NN}   % For conference papers
%\preprint{ICHEP ABSnnn}      % For conference papers
%\preprint{CLNS YY/NNNN}       % for CLNS notes
%\preprint{CLEO YY-NN}         % for CLNS notes

%\preprint{CPDRAFT2008-032}

%
\preprint{CLNS 08/2043}       % for CLNS notes
\preprint{CLEO 08-25}         % for CLNS notes

% Use this form if you DO NOT have mathematical symbols in the title
% \title{Your Title Goes Here}

% Add \boldmath if you DO have mathematical symbols in the title
%\title{\boldmath Your Title Goes Here}

\title{Improved Measurement of Absolute Branching Fraction of
$\bm{D^+_s \rightarrow \tau^+ \nu_\tau}$}

% for conference papers (ask CLEOAC for appropriate text)
%\thanks{Submitted to the 31$^{\rm st}$ International Conference on High Energy
%Physics, July 2002, Amsterdam}tmp/

%-------- INSERT HERE ------------
% Your author list goes here  REMOVE EVERYTHING to END INSERT and
% replace with your authorlist (ask cleoac).
\author{P.~U.~E.~Onyisi}
\author{J.~L.~Rosner}
\affiliation{Enrico Fermi Institute, University of
Chicago, Chicago, Illinois 60637, USA}
\author{J.~P.~Alexander}
\author{D.~G.~Cassel}
\author{J.~E.~Duboscq}\thanks{Deceased}
\author{R.~Ehrlich}
\author{L.~Fields}
\author{R.~S.~Galik}
\author{L.~Gibbons}
\author{R.~Gray}
\author{S.~W.~Gray}
\author{D.~L.~Hartill}
\author{B.~K.~Heltsley}
\author{D.~Hertz}
\author{J.~M.~Hunt}
\author{J.~Kandaswamy}
\author{D.~L.~Kreinick}
\author{V.~E.~Kuznetsov}
\author{J.~Ledoux}
\author{H.~Mahlke-Kr\"uger}
\author{D.~Mohapatra}
\author{J.~R.~Patterson}
\author{D.~Peterson}
\author{D.~Riley}
\author{A.~Ryd}
\author{A.~J.~Sadoff}
\author{X.~Shi}
\author{S.~Stroiney}
\author{W.~M.~Sun}
\author{T.~Wilksen}
\affiliation{Cornell University, Ithaca, New York 14853, USA}
% ...
\collaboration{CLEO Collaboration}
\noaffiliation

%-------- END INSERT ------------

%please hard code the date when you have a final draft and submit to CLEOAC
%\date{\today}
%
\date{January 8, 2009}

\begin{abstract} 
We have studied the leptonic decay
$D^+_s \to \tau^+ \nu_\tau$,
via the decay channel
$\tau^+ \to e^+ \nu_e \bar{\nu}_\tau$,
using a sample of tagged $D_s^+$ decays collected
near the $D^{\ast \pm}_s D^\mp_s$ peak production energy
in $e^+ e^-$ collisions with the CLEO-c detector.
%We obtain
%$\mathcal{B} (D^+_s \to \tau^+ \nu_\tau)
%= (5.30 \pm 0.47 \pm 0.22) \%$,
%where the first uncertainty is statistical and the second is systematic.
We obtain
$\mathcal{B} (D^+_s \to \tau^+ \nu_\tau)
= (5.30 \pm 0.47 \pm 0.22) \%$
and determine the decay constant
$f_{D_s} =  (252.5 \pm 11.1 \pm 5.2)$~MeV,
where the first uncertainties are statistical and the second are systematic.
\end{abstract}

\maketitle

%
% Intorduction
%    physics
%    CLEO-c
\section{\label{sec:introduction}Introduction}

The leptonic decays of a charged pseudoscalar meson $P^+$ are
processes of the type $P^+ \to \ell^+ \nu_\ell$,
where $\ell = e$, $\mu$, or $\tau$.
Because no strong interactions are present in the leptonic
final state $\ell^+ \nu_\ell$, such decays provide a clean way to
probe the complex, strong interactions that bind the quark and
antiquark within the initial-state meson. In these decays, strong
interaction effects can be parametrized by a single quantity, $f_P$,
the pseudoscalar meson decay constant.
The leptonic decay rate can be measured by experiment,
and the decay constant can be determined by the equation
(ignoring radiative corrections)
\begin{equation}
\Gamma (P_{Q\bar{q}} \rightarrow \ell^+ \nu_\ell)
=
\frac{G^2_F |V_{Qq}|^2 f^2_{P}}{8\pi}
     m_{P} m^2_\ell
     \left(1 - \frac{m^2_\ell}{m^2_{P}} \right)^2
,
\label{eq:f}
\end{equation}
where
$G_F$ is the Fermi coupling constant,
$V_{Qq}$ is the
Cabibbo-Kobayashi-Maskawa (CKM)
matrix~\cite{Cabibbo:1963yz,Kobayashi:1973fv} element,
$m_{P}$ is the mass of the meson,
and $m_\ell$ is the mass of the charged lepton.
The quantity $f_P$ describes the amplitude for the $Q$ and $\bar{q}$-quarks
within the $P$ to have zero separation, a condition necessary for
them to annihilate into the virtual $W^+$ boson that produces the
$\ell^+ \nu_\ell$ pair.

The experimental determination of decay constants is one of the
most important tests of calculations involving nonperturbative
QCD. Such calculations have been performed using various
models~\cite{Amsler:2008zz}
or using lattice QCD (LQCD). The latter is now generally considered
to be the most reliable way to calculate the quantity.

Knowledge of decay constants
is important for describing several key processes,
such as $B - \bar{B}$ mixing, which depends on $f_B$,
a quantity that is also predicted by LQCD calculations.
Experimental determination~\cite{Ikado:2006un,Aubert:2007xj}
of $f_B$ with the leptonic decay of a $B^+$ meson
is, however, very limited as the rate is highly suppressed
due to the smallness of the magnitude of the relevant CKM matrix element
$V_{ub}$.
The charm mesons, $D^+$ and $D^+_s$, are better instruments to study
the leptonic decays of heavy mesons since these decays are
either less CKM suppressed or favored, \textit{i.e.},
$\Gamma(D^+ \to \ell^+ \nu_\ell) \propto |V_{cd}|^2 \approx (0.23)^2$
and
$\Gamma(D^+_s \to \ell^+ \nu_\ell) \propto |V_{cs}|^2 \approx (0.97)^2$
are much larger than
$\Gamma(B^+ \to \ell^+ \nu_\ell) \propto |V_{ub}|^2 \approx (0.004)^2$.
Thus, the decay constants $f_D$ and $f_{D_s}$ determined
from charm meson decays can be
used to test and validate the necessary LQCD calculations
applicable to the $B$-meson sector.

Among the leptonic decays in the charm-quark sector,
$D^+_s \to \ell^+ \nu_\ell$ decays are more accessible since they are
CKM favored.
Furthermore, the large mass of the $\tau$ lepton
removes the helicity suppression that is present in the decays to
lighter leptons.
The existence of multiple neutrinos
in the final state, however, makes measurement of this decay
challenging.

Physics beyond the standard model (SM)
might also affect leptonic decays of charmed mesons.
Depending on the non-SM features, the ratio of
$\Gamma (D^+ \to \ell^+ \nu_\ell) / \Gamma (D^+_s \to \ell^+ \nu_\ell)$
could be affected~\cite{Akeroyd:2007eh},
as could
the ratio~\cite{Hewett:1995aw,Hou:1992sy}
$
\Gamma (D^+_s \to \tau^+ \nu_\tau)
/
\Gamma (D^+_s \to \mu^+ \nu_\mu)
$.
Any of the individual widths might be increased or decreased.
There is an indication of a discrepancy between
the experimental determinations~\cite{Amsler:2008zz}
of $f_{D_s}$ and the most recent
precision LQCD calculation~\cite{Follana:2007uv}.
This disagreement is particularly puzzling since the CLEO-c
determination~\cite{:2008sq}
of $f_D$ agrees well with the
LQCD calculation~\cite{Follana:2007uv}
of that quantity.
Some~\cite{Dobrescu:2008er} conjecture that
this discrepancy may be explained
by a charged Higgs boson or a leptoquark.

In this article, we report an improved
measurement of the absolute branching fraction of the
leptonic decay $D^+_s \to \tau^+ \nu_\tau$
(charge-conjugate modes are implied),
with $\tau^+ \to e^+ \nu_e \bar{\nu}_\tau$,
from which we determine the decay constant $f_{D_s}$.

\section{\label{sec:detector}Data and The CLEO-\lowercase{c} Detector}

%
% Data Sample
%
We use a
data sample of
$e^+ e^- \to D^{\ast \pm}_s D^\mp_s$
events
provided by the Cornell Electron Storage Ring (CESR)
and
collected by the CLEO-c
detector
at the center-of-mass (CM)
energy $4170$ MeV, near
$D^{\ast \pm}_s D^\mp_s$ peak production~\cite{CroninHennessy:2008yi}.
The data sample consists of an
integrated luminosity of $602$ $\text{pb}^{-1}$
containing $5.5 \times 10^5$ $D^{\ast \pm}_s D^\mp_s$ pairs.
We have previously reported~\cite{Artuso:2007zg,:2007zm}
measurements of $D^+_s \to \mu^+ \nu_\mu$ and 
$D^+_s \to \tau^+ \nu_\tau$
with a subsample of these data.
A companion article~\cite{Syracuse:2008new}
reports measurements of $f_{D_s}$ from $D^+_s \to \mu^+ \nu_\mu$
and $D^+_s \to \tau^+ \nu_\tau$, with $\tau^+ \to \pi^+ \bar{\nu}_\tau$,
using essentially the same data sample as the one  used in this measurement.  

%
% Detector
%
The CLEO-c detector~\cite{Briere:2001rn,Kubota:1991ww,cleoiiidr,cleorich}
is a general-purpose solenoidal detector with four concentric components
utilized in this measurement:  a small-radius six-layer stereo wire drift
chamber, a 47-layer main drift chamber,
a Ring-Imaging Cherenkov (RICH) detector,
and an electromagnetic calorimeter consisting of 7800 CsI(Tl) crystals.
The two drift chambers operate in a $1.0$~T magnetic field and provide
charged particle tracking in a solid angle of $93$\% of $4 \pi$.
The chambers achieve a momentum resolution of $\sim 0.6$\% at $p=1$~GeV/$c$.
The main drift chamber also provides specific-ionization ($dE/dx$)
measurements that discriminate between charged pions and kaons.
The RICH detector covers approximately $80$\% of $4 \pi$ and provides
additional separation of pions and kaons at high momentum.
The photon energy resolution of the calorimeter is $2.2$\% at
$E_\gamma=1$~GeV and $5$\% at $100$~MeV.
Electron identification is based on a likelihood variable that combines
the information from the RICH detector, $dE/dx$,
and the ratio of electromagnetic shower energy to track momentum ($E/p$).

We use a GEANT-based~\cite{geant} Monte Carlo (MC) simulation program
to study
efficiency of signal-event selection
and background processes.
Physics events are generated by \textsc{evtgen}~\cite{evtgen},
tuned with much improved knowledge of
charm decays~\cite{:2007zt,:2008cqa},
and
final-state radiation (FSR) is modeled by
the \textsc{photos}~\cite{photos} program.
The modeling of initial-state radiation (ISR) is based on cross sections
for $D^{\ast \pm}_s D^\mp_s$
production at lower energies obtained
from the CLEO-c energy scan~\cite{CroninHennessy:2008yi}
near the CM energy where we collect the sample.

%
% Analysis
%
\section{\label{sec:analysys}Analysis Method}

%
% Analysis
%
% Event Selection, Signal Selection, and Background Suppression.

%
% Analysis Overview
%
The presence of two $D^\mp_s$ mesons in a $D^{\ast \pm}_s D^\mp_s$ event
allows us to define a single-tag (ST) sample in which a $D^\mp_s$ is
reconstructed in a hadronic decay mode and a further double-tagged (DT)
subsample in which an additional $e^\pm$ is required as a
signature of $\tau^\pm$ decay, the $e^\pm$ being the daughter of
the $\tau^\pm$. The $D^-_s$ reconstructed in the ST sample can be either
primary
or secondary from $D^{\ast -}_s \to D^-_s \gamma$
(or $D^{\ast -}_{s} \to \pi^0 D^-_s$).
The ST yield can be expressed as
\begin{equation}
n_\text{ST} = 2 N \mathcal{B}_\text{ST} \epsilon_\text{ST} ,
\end{equation}
where
$N$ is the produced number of
$D^{\ast \pm}_s D^\mp_s$ pairs,
$\mathcal{B}_\text{ST}$ is the branching fraction of hadronic
modes used in the ST sample,
and $\epsilon_\text{ST}$ is the ST efficiency.
The $n_\text{ST}$ counts the candidates, not events,
and the factor of 2 comes from the sum of $D^+_s$ and $D^-_s$ tags.

Our double-tag (DT) sample is formed from events with only a single charged
track, identified as an $e^+$, in addition to a ST.
The yield
can be expressed as
\begin{equation}
n_\text{DT} =
2 N
\mathcal{B}_\text{ST}
\mathcal{B}_\text{L}
\epsilon_\text{DT},
\end{equation}
where
$\mathcal{B}_\text{L}$ is the leptonic decay branching fraction,
including the subbranching fraction of
$\tau^+ \to e^+ \nu_e \bar{\nu}_\tau$ decay,
$\epsilon_\text{DT}$ is the efficiency of finding the ST and the leptonic
decay in the same event.
From the ST and DT yields we can obtain an absolute branching fraction
of the leptonic decay $\mathcal{B}_\text{L}$,
without needing to know the integrated luminosity or the produced number of
$D^{\ast \pm}_s D^\mp_s$ pairs,
\begin{equation}
\mathcal{B}_\text{L} =
\frac{n_\text{DT}}{n_\text{ST}}
\frac{\epsilon_\text{ST}}{\epsilon_\text{DT}}
=
\frac{n_\text{DT}/\epsilon}{n_\text{ST}},
\end{equation}
where
$\epsilon$ ($\equiv \epsilon_\text{DT} / \epsilon_\text{ST}$)
is the effective signal efficiency.
Because of the large solid angle acceptance with high segmentation
of the CLEO-c detector
and
the low multiplicity of the events with which we are concerned,
$\epsilon_\text{DT} \approx \epsilon_\text{ST} \epsilon_\text{L}$,
where $\epsilon_\text{L}$ is the leptonic decay efficiency.
Hence, the ratio $\epsilon_\text{DT} / \epsilon_\text{ST}$ is insensitive
to most systematic effects associated with the ST, and the signal branching
fraction $\mathcal{B}_\text{L}$ obtained using this procedure is nearly
independent of the efficiency of the tagging mode.

\subsection{\label{sec:analysys:tag_selection}Event and tag selection}

%
% Analysis
%
% Event Selection, Signal Selection, and Background Suppression.
%
To minimize systematic uncertainties,
we tag using three two-body hadronic decay modes with only charged particles
in the final state.
The three ST modes\footnote{%
The notations $D^-_s \to \phi \pi^-$ and $D^-_s \to K^- K^{\ast 0}$ are
shorthand labels for $D^-_s \to K^- K^+ \pi^-$ events within mass windows
(described below) of the $\phi$ peak in $M(K^+K^-)$ and the $K^{\ast 0}$
peak in $M(K^+\pi^-)$, respectively. No attempt is made to separate
these resonance components in the $K^+ K^- \pi^+$ Dalitz plot.}
are
$D^-_s \to \phi \pi^-$,
$D^-_s \to K^- K^{\ast 0}$,
and
$D_s^- \to K^0_S K^-$.
Using these tag modes also helps to reduce the tag bias
which would be caused by the correlation between the tag side
and the signal side reconstruction if tag modes with
high multiplicity and large background were used.
The effect of the tag bias $b_\text{tag}$
can be expressed in terms of the signal efficiency $\epsilon$ defined by
\begin{equation}
\epsilon
=
\frac{\epsilon_\text{DT}}{\epsilon_\text{ST}}
=
\frac{\epsilon_\text{DT}}{\epsilon^\prime_\text{ST}}
\frac{\epsilon^\prime_\text{ST}}{\epsilon_\text{ST}}
=
\frac{\epsilon_\text{L} \epsilon^\prime_\text{ST}}{\epsilon^\prime_\text{ST}}
\frac{\epsilon^\prime_\text{ST}}{\epsilon_\text{ST}}
=
\epsilon_\text{L} b_\text{tag}
,
\end{equation}
where
$\epsilon^\prime_\text{ST}$ is the ST efficiency when the recoiling
system is the signal leptonic decay with single $e^\pm$ in the other side
of the tag.
As the general ST efficiency
$\epsilon_\text{ST}$, when the recoiling system is any possible $D_s$ decays,
will be lower than the $\epsilon^\prime_\text{ST}$, sizable
tag bias could be introduced if the multiplicity of the tag mode
were high, or the tag mode were to include neutral particles
in the final state.
As shown in  Sec.~\ref{sec:results},
this effect is negligible in our chosen clean tag modes.

The $K^0_S \to \pi^+ \pi^-$ decay is reconstructed by combining
oppositely charged tracks that originate from a common vertex
and that have an invariant mass within $\pm 12$ MeV of the
nominal mass~\cite{Amsler:2008zz}.
We require the
resonance decay to satisfy the following mass windows around the
nominal masses~\cite{Amsler:2008zz}:
$\phi \to K^+ K^-$ ($\pm 10$ MeV)
and
$K^{\ast 0} \to K^+ \pi^-$ ($\pm 75$ MeV).
We require the momenta of charged particles to be
$100$ MeV or greater to suppress the slow pion background from
$D^\ast \bar{D}^\ast$ decays (through $D^\ast \to \pi D$).
We identify a ST by using
the invariant mass of the tag $M(D_s)$
and recoil mass against the tag $M_\text{recoil}(D_s)$.
The recoil mass is defined as
\begin{equation}
M_\text{recoil}(D_s)
\equiv
\sqrt{ (E_{ee} - E_{D_s} )^2 - |{\bf p}_{ee} - {\bf p}_{D_s}|^2 }
,
\end{equation}
where $(E_{ee}, {\bf p}_{ee})$ is the net four-momentum of the $e^+ e^-$ beam,
taking the finite beam crossing angle into account;
$(E_{D_s},{\bf p}_{D_s})$ is the four-momentum of the tag,
with $E_{D_s}$ computed from ${\bf p}_{D_s}$ and
the nominal mass~\cite{Amsler:2008zz} 
of the $D_s$ meson.
We require the recoil mass to be
within $55$ MeV of the $D^\ast_s$ mass~\cite{Amsler:2008zz}.
This loose window allows both
primary and secondary $D_s$ tags to be selected.

\begin{figure*}
\centering
\includegraphics[width=0.95\textwidth]{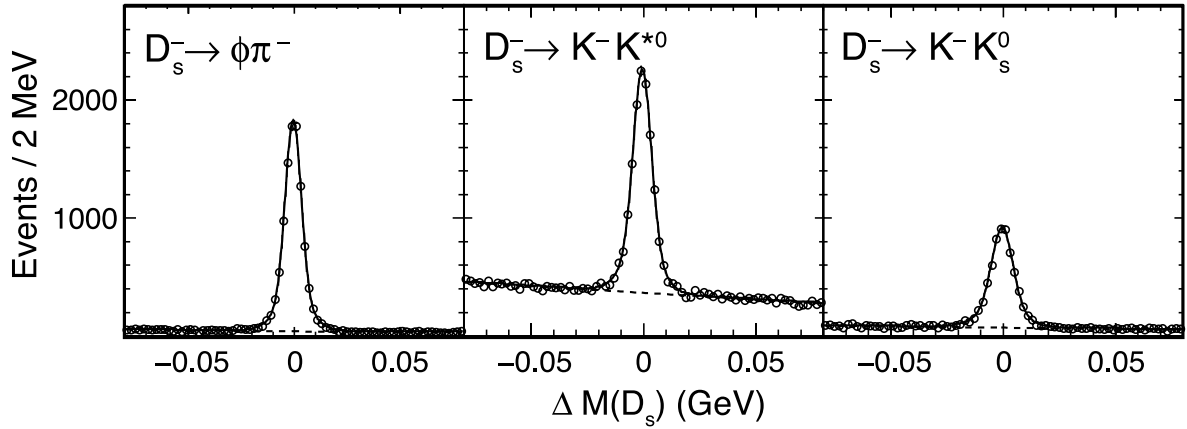}
\caption{\label{fig:dm}
The mass difference $\Delta M (D_s) \equiv M(D_s) - m_{D_s}$
distributions in each tag mode.
We fit the $\Delta M(D_s)$
%distribution (filled circle)
%
distribution (open circle)
to the sum (solid curve)
of signal function (double Gaussian)
plus background function (second-degree Chebyshev polynomial, dashed curve).
}
\end{figure*}

To estimate the backgrounds
in our ST and DT yields from the wrong tag combinations
(incorrect combinations that, by chance, lie within the
$\Delta M(D_s)$ signal region),
we use the tag invariant mass sidebands.
We define the signal region as 
$-20$ MeV $\le \Delta M(D_s) < +20$ MeV, and the
sideband regions as
$-55$ MeV $\le \Delta M(D_s) < -35$ MeV
or  $+35$ MeV $\le \Delta M(D_s) < +55$ MeV,
where $\Delta M(D_s) \equiv M(D_s) - m_{D_s}$
is the difference between the tag mass and the nominal mass.
We fit the ST $\Delta M(D_s)$ distributions
to the sum of double-Gaussian signal function
plus second-degree Chebyshev polynomial
background function to get the tag mass sideband scaling factor.
The invariant mass distributions of tag candidates
for each tag mode are shown in Fig.~\ref{fig:dm}
and the ST yield and $\Delta M(D_s)$ sideband scaling factor
are summarized in Table~\ref{table:data-single}.
We find $n_\text{ST} = 26334 \pm 213$
summed over the three tag modes.

\begin{table}
\centering
\caption{\label{table:data-single}
Summary of single-tag (ST) yields, where $n^\text{S}_\text{ST}$ is
the yield in the ST mass signal region,
$n^\text{B}_\text{ST}$ is the yield in the sideband region,
$s$ is the sideband scaling factor,
and
$n_\text{ST}$ is the scaled sideband-subtracted yield.
}
\begin{ruledtabular}
\begin{tabular}{lrrrr}
Tag mode & $n^\text{S}_\text{ST}$& $n^\text{B}_\text{ST}$& $s$& $n_\text{ST}$\\
\hline
$D_s^- \rightarrow \phi \pi^-$ &
$10459$& $807$& $0.980$& $9668.1 \pm 106.1$\\
$D_s^- \rightarrow K^- K^{\ast 0}$ &
$18319$& $7381$& $1.000$& $10938.0 \pm 160.3$\\
$D_s^- \rightarrow K^- K^0_S$ &
$7135$& $1409$& $0.999$& $5727.8 \pm 92.4$\\
\hline
\multicolumn{4}{l}{Total} &
$26333.9 \pm 213.3$\\
\end{tabular}
\end{ruledtabular}
\end{table}

\subsection{\label{sec:analysys:signal_selection}Signal-event selection}

A DT event is required to have a ST,
a single $e^+$, no additional charged particles, and the
net charge of the event $Q_\text{net} = 0$.
We require the momentum of the $e^+$ candidate be at least
$200$~MeV.

The DT events will contain the sought-after
$D^+_s \to \tau^+ \nu_\tau$ ($\tau^+ \to e^+ \nu_e \bar{\nu}_\tau$) events,
but also some backgrounds.
The most effective variable for separating
signal from background events is the extra energy ($E_\text{extra}$)
in the event, \textit{i.e.}, the total energy of the rest of the event
measured in the electromagnetic calorimeter.
This quantity is computed
using the neutral shower energy in the calorimeter, counting all
neutral clusters consistent with being photons above $30$ MeV;
these showers must not be associated with any of the ST decay
tracks or the signal $e^+$.
We obtain $E_\text{extra}$ in the signal and sideband regions
of $\Delta M(D_s)$. The sideband-subtracted $E_\text{extra}$
distribution is used to obtain the DT yield.

\begin{figure}
\centering
\includegraphics[width=\columnwidth]{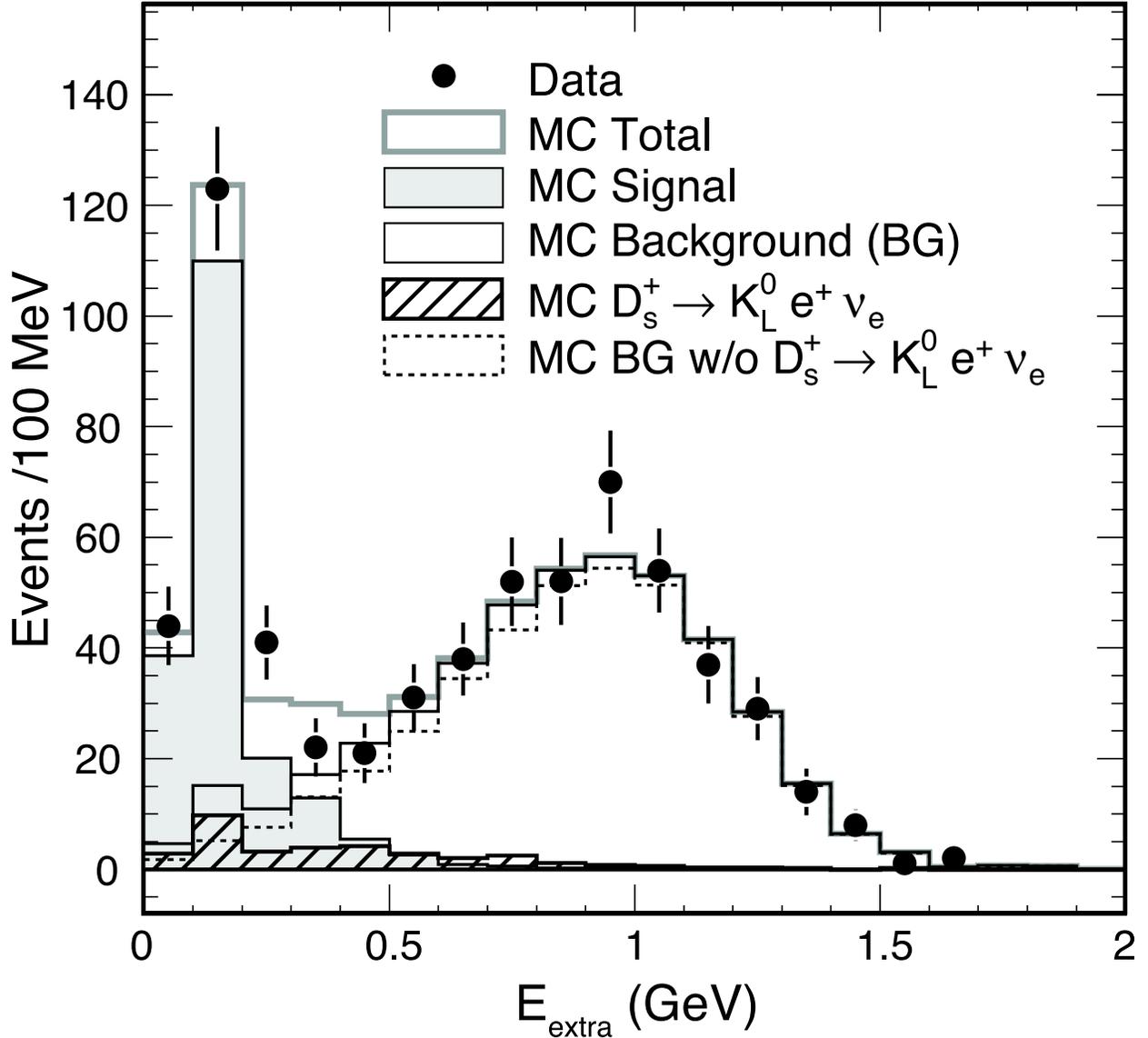}
\caption{\label{fig:ecc}
Distribution of $E_\text{extra}$ after $\Delta M(D_s)$
sideband subtraction.
Filled circles are from data and
histograms are obtained from MC simulation.
The MC signal and peaking background ($D^+_s \rightarrow K^0_L e^+ \nu_e$)
components are normalized to our measured branching fractions.
The errors shown are statistical only.
}
\end{figure}

The $E_\text{extra}$ distribution obtained from data is
compared to the MC expectation in Fig.~\ref{fig:ecc}.
We have used the invariant mass
sidebands, defined in Sec.~\ref{sec:analysys:tag_selection},
to subtract the combinatorial background.
We expect that there will be a large peak between
$100$ MeV and $200$ MeV from $D_s^*\to \gamma D_s$ decays
(and from $D_s^* \to \pi^0 D_s$,
with a $5.8 \%$ branching fraction~\cite{Amsler:2008zz}).
Also, there will be some events at lower energy when the photon
from $D_s^*$ decay escapes detection.
Based on considerations described in the next paragraph,
we define our signal region to be $E_\text{extra} < 400$ MeV.

\subsection{\label{sec:analysys:background}Background estimation}

After the $\Delta M(D_s)$ sideband subtraction, two significant
components of background remain.
One is from $D^+_s \rightarrow K^0_L e^+ \nu_e$ decay.
If the $K^0_L$ deposits
little or no energy in the calorimeter, this decay
mode has an $E_\text{extra}$ distribution very similar to the signal,
peaking well below $400$ MeV.
The second source, other $D_s$ semielectronic decays,
rises smoothly with increasing $E_\text{extra}$, up to $1$ GeV.
Estimates of these backgrounds
are also shown in Fig.~\ref{fig:ecc}.
The optimal signal region in $E_\text{extra}$ for DT yield extraction
is predicted from an MC simulation study.
Choosing $E_\text{extra}$ less than $400$ MeV
maximizes the signal significance.
Note that with our
chosen requirement of  $E_\text{extra} < 400$ MeV, we are including
$D^+_s \to \tau^+ \nu_\tau \gamma$ as signal.
However, this is expected to be very small,
as the kinetic energy of the $\tau^+$ in the $D^+_s$ rest frame is
only $9.3$ MeV and it cannot radiate much.

The number of nonpeaking background events $b_\text{np}$
in the $E_\text{extra}$ signal region
is estimated from the number of events
in the $E_\text{extra}$ sideband region
between $0.6$~GeV and $2$~GeV,
scaled by the MC-determined ratio $c_b$ ($\equiv b^\text{(l)} / b^\text{(h)}$)
of the number of background events
in the $E_\text{extra}$ signal region, $b^\text{(l)}$,
to the number of events
in the $E_\text{extra}$ sideband region, $b^\text{(h)}$.
The number of peaking background events $b_\text{p}$
due to the
$D^+_s \rightarrow K^0_L e^+ \nu_e$ decay
is determined by using the expected number from MC simulation.
The overall expected number of background events
in the $E_\text{extra}$ signal
region, $b$, is computed as follows:
\begin{equation}
b
=
b_\text{np} + b_\text{p}
=
c_{b} b^\text{(h)}(\text{data})
+
b(K^0_L e^+ \nu_e)_\text{MC}
,
\end{equation}
where
$b^\text{(h)}(\text{data})$ is the number of data events
in the $E_\text{extra}$ sideband region and
$b(K^0_L e^+ \nu_e)_\text{MC}$ is the number of background events
due to $D^+_s \rightarrow K^0_L e^+ \nu_e$ as estimated
from our MC simulation.
We normalize this quantity using our measured~\cite{koloina}
$\mathcal{B} (D^+_s \rightarrow K^0_S e^+ \nu_e)
= (0.19 \pm 0.05 \pm 0.01) \%$.
We simulate calorimeter response to $K^0_L$
using a momentum dependent
$K^0_L$ interaction probability density function obtained from
studying
$\psi (3770) \to D^0 \bar{D}^0$ events in which the
$\bar{D}^0$ has been reconstructed in hadronic tag modes and the $D^0$
decays to the $K^0_L \pi^+ \pi^-$ final state.

The numbers of estimated background events from peaking
and nonpeaking sources in each tag mode are summarized
in Table~\ref{table:background}.

\begin{table}
\centering
\caption{\label{table:background}Estimated backgrounds in the extra
energy signal region below $400$ MeV in each tag mode.  Here
$b_\text{p}$ is the peaking background from
$D^+_s \to K^0_L e^+ \nu_e$ decay,
$b_\text{np}$ is the nonpeaking background from other
$D_s$ semileptonic decays,
and $b$ is the total number of background events.
The errors shown are statistical only.}
\begin{ruledtabular}
\begin{tabular}{lrrr}
Tag mode & $b_\text{np}$& $b_\text{p}$& $b$\\
\hline
$D_s^- \rightarrow \phi \pi^-$& $11.8 \pm 1.0$& $7.7 \pm 0.5$& $19.4 \pm 1.1$\\
$D_s^- \rightarrow K^- K^{\ast 0}$& $12.2 \pm 1.1$& $8.7 \pm 0.5$& $20.9 \pm 1.3$\\
$D_s^- \rightarrow K^- K^0_S$& $4.6 \pm 0.7$& $4.5 \pm 0.3$& $9.1 \pm 0.7$\\
\hline
Total &
$28.5 \pm 1.7$ & $20.9 \pm 0.8$ & $49.4 \pm 1.8$\\
\end{tabular}
\end{ruledtabular}
\end{table}

%
% Results
%
\section{\label{sec:results}Results}

The signal efficiency determined by MC simulation has been
weighted by the ST yields in each mode as shown
in Table~\ref{table:efficiency}.
We determine the weighted average signal efficiency
$\epsilon = (72.4 \pm 0.3) \%$
for the decay chain
$D^+_s \to \tau^+ \nu_\tau \to e^+ \nu_e \bar{\nu}_\tau  \nu_\tau$.

\begin{table}
\centering
\caption{\label{table:efficiency}
Summary of the signal efficiency determined by MC simulation.
Average efficiency $\epsilon$ and the tag bias $b_\text{tag}$ are obtained
by using the weighting factor $w$ determined from
single-tag yields in data.}
\begin{ruledtabular}
\begin{tabular}{lrrr}
Tag mode & $w$& $\epsilon \equiv \epsilon_\text{DT} / \epsilon_\text{ST}$ & $b_\text{tag} = \epsilon^\prime_\text{ST} / \epsilon_\text{ST}$ \\
\hline
$D_s^- \rightarrow \phi \pi^-$&
$0.3671$& 
$0.6964 \pm 0.0046$& 
$1.0089 \pm 0.0058$
\\
$D_s^- \rightarrow K^- K^{\ast 0}$&
$0.4154$& 
$0.7337 \pm 0.0049$& 
$1.0061 \pm 0.0060$
\\
$D_s^- \rightarrow K^- K^0_S$&
$0.2175$& 
$0.7536 \pm 0.0054$& 
$1.0032 \pm 0.0065$
\\
\hline
Average & 
 & 
$0.7244 \pm 0.0029$& 
$1.0065 \pm 0.0036$
\\
\end{tabular}
\end{ruledtabular}
\end{table}

The DT yields with the $400$ MeV extra energy requirement
are summarized in Table~\ref{table:data-double}.
We find $n_\text{DT} = 180.6 \pm 15.9$ summed over all tag modes.
Using
$\mathcal{B} (\tau^+ \rightarrow e^+ \nu_e \bar{\nu}_\tau)
= (17.85 \pm 0.05) \%$~\cite{Amsler:2008zz},
we obtain the leptonic decay branching fraction
$\mathcal{B} (D^+_s \rightarrow \tau^+ \nu_\tau)
= (5.30 \pm 0.47) \%$, where the uncertainty is statistical.

\begin{table}
\centering
\caption{\label{table:data-double}
Summary of double-tag (DT) yields in each tag mode,
where
$n^\text{S}_\text{DT}$ is the DT yield in the tag mass signal region,
$n^\text{B}_\text{DT}$ is the yield in the tag mass sideband region,
$s$ is the tag mass sideband scaling factor,
$b$ is the number of estimated background in the extra energy signal
region after tag mass sideband scaled background subtraction,
and
$n_\text{DT}$ is the background subtracted DT yield.
The errors shown are statistical only.}
\begin{ruledtabular}
\begin{tabular}{lrrrrr}
Tag mode & $n^\text{S}_\text{DT}$& $n^\text{B}_\text{DT}$& $s$& $b$& $n_\text{DT}$\\
\hline
$D_s^- \rightarrow \phi \pi^-$& $79$& $1$& $0.980$& $19.4 \pm 1.1$& $58.6 \pm 9.0$\\
$D_s^- \rightarrow K^- K^{\ast 0}$& $110$& $6$& $1.000$& $20.9 \pm 1.3$& $83.1 \pm 10.8$\\
$D_s^- \rightarrow K^- K^0_S$& $50$& $2$& $0.999$& $9.1 \pm 0.7$& $38.9 \pm 7.2$\\
\hline
\multicolumn{4}{l}{Total} &
$49.4 \pm 1.8$& $180.6 \pm 15.9$\\
\end{tabular}
\end{ruledtabular}
\end{table}

%
% Systematic uncertainty
%
\section{\label{sec:systematic_uncertainty}Systematic Uncertainty}

Sources of systematic uncertainties and their effects on the
$D^+_s \to \tau^+ \nu_\tau$
branching fraction determination
are summarized in Table~\ref{table:systematics}.

\begin{table}
\centering
\caption{\label{table:systematics}Summary of sources of systematic
uncertainty and their effects on the branching fraction measurement.
}
\begin{ruledtabular}
\begin{tabular}{lr}
Source &
Effect on $\mathcal{B}$ $(\%)$
\\
\hline
Background (nonpeaking)&
$0.7$
\\
$D^+_s \to K^0_L e^+ \nu_e$ (peaking)&
$3.2$
\\
Extra shower &
$1.1$
\\
Extra track &
$1.1$
\\
$Q_\text{net} = 0$ &
$1.1$
\\
Non electron&
$0.1$
\\
Secondary electron&
$0.3$
\\
Number of tag&
$0.4$
\\
Tag bias&
$0.2$
\\
Tracking&
$0.3$
\\
Electron identification&
$1.0$
\\
FSR&
$1.0$
\\
\hline
Total&
$4.1$
\end{tabular}
\end{ruledtabular}
\end{table}

%
% Systematic uncertainty
%
We considered six semileptonic decays,
$D^+_s \to$
$\phi e^+ \nu_e$,
$\eta e^+ \nu_e$,
$\eta' e^+ \nu_e$,
$K^0 e^+ \nu_e$,
$K^{\ast 0} e^+ \nu_e$,
and
$f_0 e^+ \nu_e$,
as the major sources of background in the $E_\text{extra}$ signal
region. The second dominates the nonpeaking background,
and the fourth (with $K^0_L$) dominates the peaking background.
Uncertainty in the signal yield due to nonpeaking background ($0.7\%$)
is assessed by varying the semileptonic decay branching fractions
by the precision with which they are known~\cite{koloina}.
Imperfect knowledge of $\mathcal{B}(D^+_s \to K^0 e^+ \nu_e)$
gives rise to a systematic uncertainty in our estimate of the amount
of peaking background in the signal region, which has an effect on our
branching fraction measurement of $3.2 \%$.

We study differences in efficiency, data vs  MC events, due to
the extra energy requirement, extra track veto,
and $Q_\text{net} = 0$ requirement, by using samples from data and
MC events, in which \textit{both} the $D_s^-$ and $D_s^+$
satisfy our tag requirements, i.e., ``double-tag'' events.
We then apply each of the above-mentioned requirements and
compare loss in efficiency of data vs  MC events. In this way 
we obtain a correction of $1.6\%$ for the extra energy
requirement and systematic uncertainties on each of the three
requirements of $1.1\%$ (all equal, by chance).

The non-$e^+$ background in the signal
$e^+$ candidate sample is negligible ($0.4\%$)
due to the low probability ($\sim 0.1\%$ per track)
that hadrons ($\pi^+$ or $K^+$) are misidentified as $e^+$~\cite{Adam:2006nu}.
Uncertainty in these backgrounds produces a
$0.1\%$ uncertainty in the measurement
of $\mathcal{B} (D^+_s \rightarrow \tau^+ \nu_\tau)$.
The secondary $e^+$ backgrounds
from charge symmetric processes,
such as $\pi^0$ Dalitz decay ($\pi^0 \to e^+ e^- \gamma$)
and $\gamma$ conversion ($\gamma \to e^+ e^-$), are assessed
by measuring the wrong-sign signal electron in events with
$Q_\text{net} = \pm 2$.
The uncertainty in the measurement from this source is estimated to be
$0.3\%$.

Other possible sources of systematic uncertainty
include
$n_\text{ST}$ ($0.4\%$),
tag bias ($0.2 \%$),
tracking efficiency ($0.3\%$),
$e^\pm$ identification efficiency ($1\%$),
and FSR ($1\%$).
Combining all contributions in quadrature,
the total systematic uncertainty in the branching fraction measurement
is estimated to be
$4.1 \%$.

%
% SUMMARY
%
\section{\label{sec:conclusion}Summary}

In summary, using the sample of $26\, 334$ 
tagged $D^+_s$ decays with the CLEO-c detector
we obtain the absolute branching fraction of the leptonic decay
$D^+_s \rightarrow \tau^+ \nu_\tau $ through
$\tau^+ \rightarrow e^+ \nu_e \bar{\nu}_\tau$
\begin{equation}
\mathcal{B} (D^+_s \rightarrow \tau^+ \nu_\tau)
=  (5.30 \pm 0.47 \pm 0.22) \%
,
\end{equation}
where the first uncertainty is statistical and the second is systematic.
This result supersedes our previous measurement~\cite{:2007zm}
of the same branching fraction,
which used a subsample of data used in this work.

The decay constant $f_{D_s}$
can be computed using Eq.~(\ref{eq:f})
with known values~\cite{Amsler:2008zz}
$G_F = 1.16637(1)\times 10^{-5}$ GeV$^{-2}$,
$m_{D_s} = 1968.49(34)$ MeV,
$m_\tau = 1776.84(17)$ MeV,
and $\tau_{D_s} = 500(7) \times 10^{-15}$ s.
We assume $|V_{cs}| = |V_{ud}|$
and use the value $0.97418(26)$ given in Ref.~\cite{Towner:2007np}.
We obtain
\begin{equation}
f_{D_s} =  (252.5 \pm 11.1 \pm 5.2) \text{MeV}.
\end{equation}

Combining with our other determination~\cite{Syracuse:2008new}
of $f_{D_s} = (263.3 \pm 8.2 \pm 3.9)$ MeV
with 
$D^+_s \to \mu^+ \nu_\mu$
and
$D^+_s \to \tau^+ \nu_\tau$ ($\tau^+ \to \pi^+ \bar{\nu_\tau}$)
decays,
we obtain
\begin{equation}
f_{D_s} = (259.5 \pm 6.6 \pm 3.1) \text{MeV}.
\end{equation}
This result is derived from absolute branching fractions only
and is the most precise determination of the $D_s$ leptonic decay
constant to date.

Our combined result is larger than the recent LQCD calculation
$f_{D_s} = (241 \pm 3)$ MeV~\cite{Follana:2007uv}
by $2.3$ standard deviations.
The difference between data and LQCD for $f_{D_s}$ could be due to
physics beyond the SM~\cite{Dobrescu:2008er},
unlikely statistical fluctuations
in the experimental measurements or the LQCD calculation,
or systematic uncertainties that are not understood in the LQCD calculation
or the experimental measurements.

Combining with our other determination~\cite{Syracuse:2008new}
of
$\mathcal{B}(D^+_s \rightarrow \tau^+ \nu_\tau) = (6.42 \pm 0.81 \pm 0.18) \%$,
via $\tau^+ \to \pi^+ \bar{\nu}_\tau$,
we obtain
\begin{equation}
\mathcal{B} (D^+_s \rightarrow \tau^+ \nu_\tau)
  =  (5.62 \pm 0.41 \pm 0.16) \%.
\end{equation}
Using this with our measurement~\cite{Syracuse:2008new}
of
$\mathcal{B}(D^+_s \rightarrow \mu^+ \nu_\mu) = (0.565 \pm 0.045 \pm 0.017) \%$,
we obtain the branching fraction ratio
\begin{equation}
R
=
\frac{\mathcal{B}(D^+_s \rightarrow \tau^+ \nu_\tau)}
     {\mathcal{B}(D^+_s \to \mu^+ \nu_\mu)}
=
10.1 \pm 0.9 \pm 0.3.
\end{equation}
This is consistent with $9.76$, the value predicted by the SM
with lepton universality,
as given in Eq.~(\ref{eq:f}) with known masses~\cite{Amsler:2008zz}.

\begin{acknowledgments}
We gratefully acknowledge the effort of the CESR staff
in providing us with excellent luminosity and running conditions.
D.~Cronin-Hennessy and A.~Ryd thank the A.P.~Sloan Foundation.
This work was supported by the National Science Foundation,
the U.S. Department of Energy,
the Natural Sciences and Engineering Research Council of Canada, and
the U.K. Science and Technology Facilities Council.
\end{acknowledgments}

\end{document}